\documentclass[preprint,12pt]{elsarticle}

\usepackage{graphics}
\usepackage{graphicx}

\usepackage{amsmath}
\usepackage{amsthm}
\usepackage{epstopdf}
\usepackage{dcolumn}
\usepackage{bm}
\usepackage{color}
\usepackage[
total={6.5in,8.75in}, top=1.5in, left=0.9in, 
]{geometry}

\biboptions{comma,sort,compress}

\journal{Physica A}

\begin{document}

\begin{frontmatter}



\title{Income and wealth distribution of the richest Norwegian
  individuals: An inequality analysis} 

\author[1,2,3]{Maciej Jagielski\corref{cor1}}
\ead{zagielski@interia.pl}
\cortext[cor1]{Corresponding author. Tel.: +48 22 5532730; fax: +48 22
  5532999.} 
\author[1]{Kordian Czy\.zewski}
\author[1]{Ryszard Kutner}
\author[3]{H. Eugene Stanley}

\address[1]{Faculty of Physics, University of Warsaw, Pasteura 5,
  PL-02093 Warszawa, Poland} 
\address[2]{ETHZ, Department of Management, Technology and Economics,
  Scheuchzerstrasse 7, CH-8092 Z\"urich, Switzerland} 
\address[3]{Center for Polymer Studies and Department of Physics, Boston
  University, Boston, MA 02215 USA} 

\begin{abstract}

Using the empirical data from the Norwegian tax office, we analyse the
wealth and income of the richest individuals in Norway during the period
2010--2013. We find that both annual income and wealth level of the
richest individuals are describable using the Pareto law. We find that
the robust mean Pareto exponent over the four-year period to be $\approx
2.3$ for income and $\approx 1.5$ for wealth.

\end{abstract}


\end{frontmatter}



\section{Introduction}\label{intro}

Income and wealth inequalities are being closely examined in current
economic, sociological and econophysical literature \cite{P_2014,
  AB_2015, CG_2005a, CGK_2009, DY_2001a, JKP_2012, L_1998, L_2003,
  NS_2007, YR_2009, JDK_2015, JK_2013b, SPW_2004b, CMG_2006, CG_2005b,
  P_1897, ASF_2003,ASNOTT_2000,FSAKA_2003,S_2001,MAH_2004,
  RHCR_2006,S_2006,KY_2004,DNS_2012, JK_2010,KBLMS_2007,LS_1997}. The
challenge is to accurately measure these inequalities.

The recent, widely-cited book on income and wealth inequalities by
Piketty \cite{P_2014} concludes that income and wealth inequalities are
different quantities and should be analyzed separately.  Many authors
have used the Pareto law to describe income or/and wealth inequalities
\cite{CG_2005a, CGK_2009, DY_2001a, JKP_2012, L_1998, L_2003, NS_2007,
  YR_2009, JDK_2015, JK_2013b, SPW_2004b, CMG_2006, CG_2005b, P_1897,
  ASF_2003,ASNOTT_2000,FSAKA_2003,S_2001,MAH_2004,
  RHCR_2006,S_2006,KY_2004,DNS_2012, JK_2010,KBLMS_2007,LS_1997}.
Piketty uses aggregated macro-variables to describe inequality, but
these authors following the Pareto approach primarily use microdata,
i.e., the wealth ranks of the richest individuals supplied by such
periodicals as {\it Forbes}. In both kinds of analysis the quality of
the empirical data is poor. Piketty's empirical data, although reliable,
are imperfect, and those used by other researchers are also far from
perfect, because the only reliable source for such data is the tax
office.

Since the publication of Vilfredo Pareto's pioneering work on income
distribution in the late nineteenth century many additional studies have
been carried out to empirically verify the Pareto law for both
individuals and households. These analyses were carried out for the
United States \cite{CG_2005a, CGK_2009,DY_2001a,JKP_2012,
  L_1998,L_2003,NS_2007,YR_2009,JDK_2015}, the European Union
\cite{JKP_2012, JK_2013b, JDK_2015}, the UK
\cite{CG_2005a,DY_2001a,L_1998,L_2003, SPW_2004b}, Germany
\cite{CG_2005a}, Italy \cite{CMG_2006,CG_2005b}, France \cite{L_1998},
Switzerland \cite{P_1897}, Sweden \cite{L_2003}, Japan
\cite{ASF_2003,ASNOTT_2000,FSAKA_2003,NS_2007,S_2001}, Australia
\cite{CMG_2006,MAH_2004}, Canada \cite{RHCR_2006}, India
\cite{RHCR_2006,S_2006}, Sri Lanka \cite{RHCR_2006}, Peru \cite{P_1897},
Egypt \cite{RHCR_2006}, South Korea \cite{KY_2004}, Romania
\cite{DNS_2012}, Portugal \cite{RHCR_2006}, Poland
\cite{JK_2010,JKP_2012}, and for the world as a whole
\cite{KBLMS_2007,LS_1997}.

Here we use the Pareto law to analyze the income and wealth rank of the
100 richest individuals in Norway. Beginning in the nineteenth century,
the tax office in Norway has compiled a yearly ``Skattelister,'' a list
of the yearly income and wealth level of every citizen in
Norway. Although access to the Skattelister has been extremely limited,
the records of the 100 richest individuals are publicly available for
the years 2010--2013. This allows a precise validation of the Pareto law
for income and wealth distributions during those years
\cite{NTO_2010_2013}. Note that although current literature provides
numerous occasions of this type of analysis, the Skattelister data we
obtained from the Norwegian tax office makes our validation particularly
reliable.

\section{Pareto law}

At the end of nineteenth century Vilfredo Pareto formulated his law by
analyzing a huge amount of empirical data that described the income and
wealth distributions using the PDF universal function, i.e.
\cite{P_1897, M_1960, JK_2013b},
\begin{equation}
P(m)=\left\{\begin{array}{cc}\displaystyle
\frac{\alpha}{m_{0}}\left(\frac{m_{0}}{m}\right)^{\alpha+1} &
\mathrm{if} \quad m \ge m_{0}  \\  0 & \mathrm{if} \quad
m<m_{0}\end{array}\right. , 
\label{rown1}
\end{equation}
which defines the Pareto law \cite{M_1960}. The $m_0$ value is the
lowest value of variable $m$ and $\alpha$ is the Pareto exponent. By
definition we choose the strongest Pareto law \cite{M_1960} for
$\alpha=3/2$.

Empirical studies indicate (i) that the mean value of the Pareto
exponent is close to $2$, and (ii) that the Pareto law is valid for
large values of income and wealth \cite{RHCR_2006}. For other values of
income and wealth such laws as Gibrat's rule of proportionate growth
\cite{RHCR_2006} are valid. Thus the studies in Section 1 refer to the
weak Pareto law that holds in the limit $m \gg m_0$. We will show that
the mean value of the Pareto exponent for the richest Norwegians is
close to $2$ for income and to $3/2$ for wealth.

To analyse the empirical data it is better to use the more robust
(global) empirical complementary cumulative distribution function (CCDF)
rather than the (local) Eq.~(\ref{rown1}). From Eq.~(\ref{rown1}) the
CCDF can be written
\begin{equation}
\Pi(m) = (m/m_{0})^{-\alpha}, \quad \textrm{for }m\geq m_0.
\label{rown2}
\end{equation}

A well-known feature of Eq.~\ref{rown1} is the divergence of its moments
of an order greater than or equal to $\alpha$. This means that we should
apply quantiles (which are always finite \cite{HF_1996}) instead of
moments. For example, we use the median instead of the mean value. In
any case, the moment estimates are always finite and can be calculated
directly from the empirical data \cite{JK_2013b}.

\section{Results and concluding remarks}

The annual rankings of income and wealth of Norwegian citizens for the
years 2010--2013 \cite{NTO_2010_2013} are of the 100 wealthiest people
in Norway as well as in each region (fylke) of Norway. Note that income
and wealth ranking lists are reported independently. Someone listed in
the income ranking is not listed in the wealth ranking and vice
versa. Unlike those in, for example, {\it Forbes}, the data supplied by
the Norwegian tax office are not estimates and are of the highest
quality.

\begin{figure}[ht]
\centering 
\includegraphics[scale=0.505]{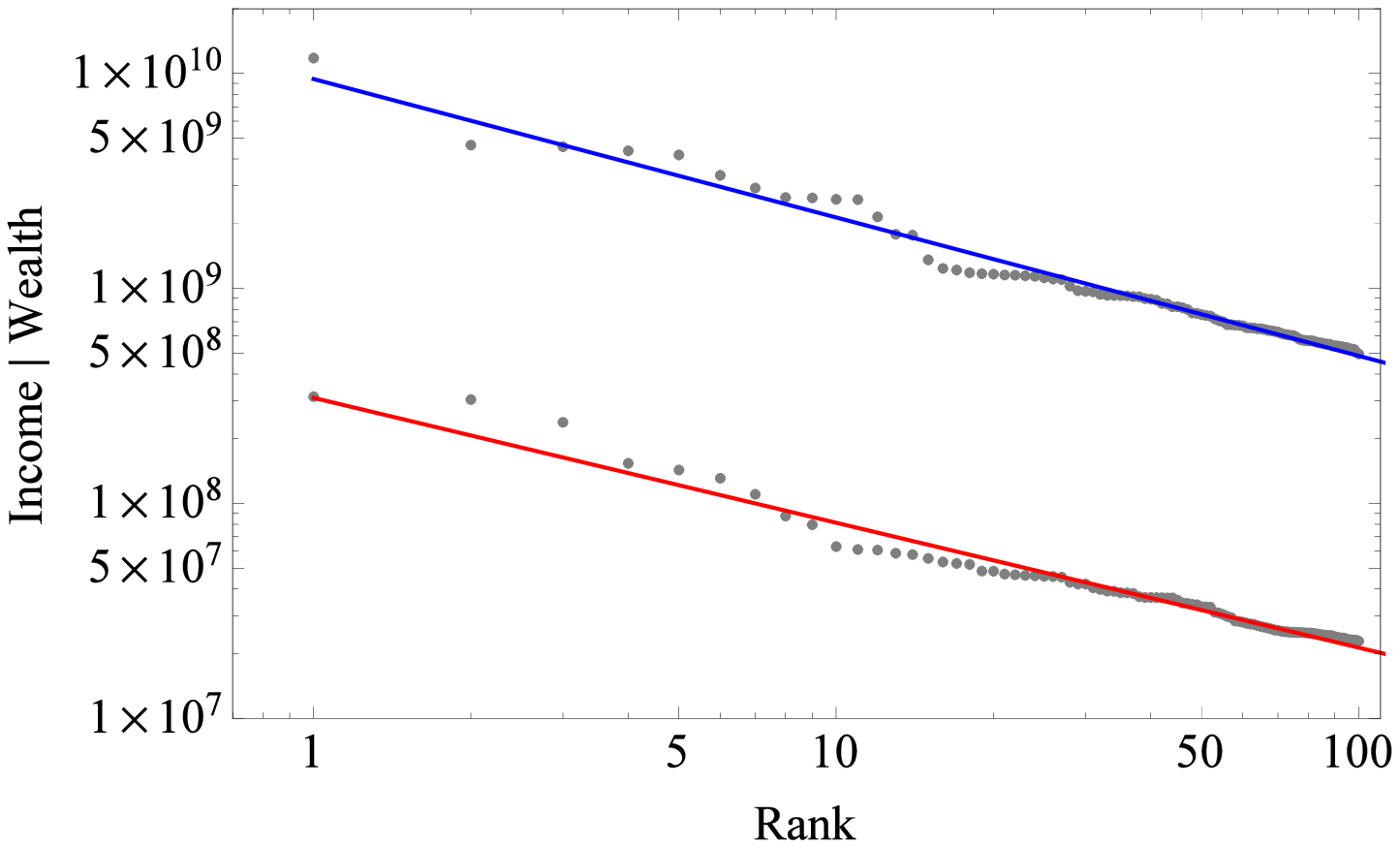}
\includegraphics[scale=0.5]{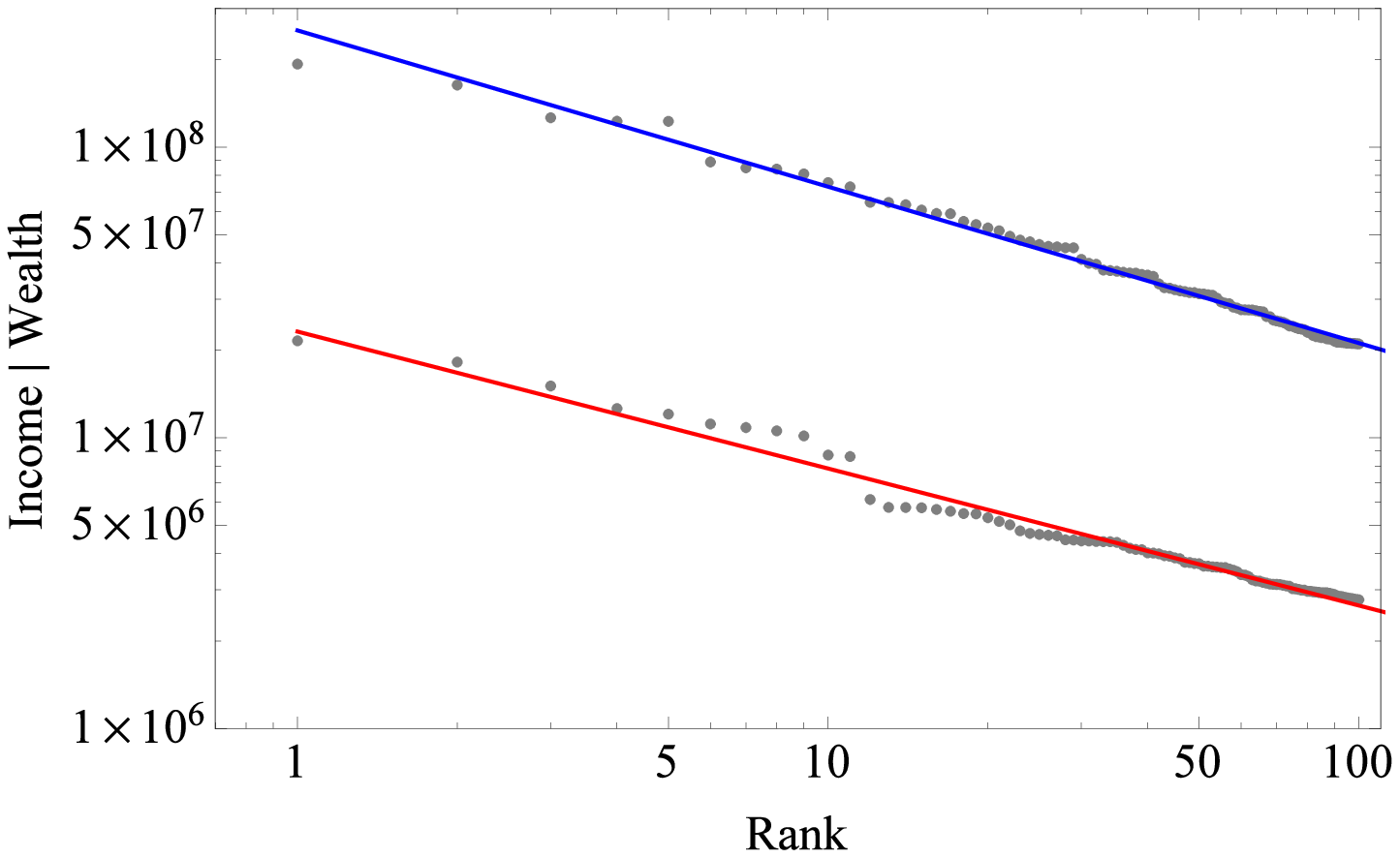}
\caption{Ranks of income (bottom plots) and wealth (top plots) of the
  richest individuals in Norway (left panel) and Hedmark region (right
  panel). Solid lines were obtained by fitting straight lines (in the
  log-log scale) to empirical data (dots) for year 2013.}
\label{fig1}
\end{figure}

Figure \ref{fig1} shows log-log plots of the wealth and income rankings
for the Hedmark region and Norway as a whole for the year 2013.  The
slopes of these lines are equal to $-\alpha_{\textrm{rank}}$ and were
calculated using a fitting routine. The inverse of
$\alpha_{\textrm{rank}}$ gives the Pareto exponent $\alpha $.

Table \ref{tab1} shows the Pareto exponents for each region of Norway
and for all of Norway for the years 2010--2013. The error bars of the
Pareto exponents fall within the range 0.01--0.13.

\begin{table}[ht]
\centering
\caption{Pareto exponents for each region of Norway and all of Norway
  the for years 2010--2013. The errors bars of Pareto exponents fall within the
  range $0.01-0.13$.}
\begin{tabular}{|l||l|l|l|l|l|l|l|l|l|} \hline
Region/year & \multicolumn{2}{|c|}{2010} & \multicolumn{2}{|c|}{2011} & \multicolumn{2}{|c|}{2012} & \multicolumn{2}{|c|}{2013}  \\ \hline
& income & wealth & income & wealth & income & wealth & income & wealth \\ \hline \hline
{\small Akershus} & $2.45$ & 1.14 & 3.00 & 1.21 & 2.34 & 1.23 & 2.45 & 1.43\\ \hline
{\small Aust-Agder} & $2.22$ & 1.20 & 2.19 & 1.13 & 1.70 & 1.14 & 1.58 & 1.16\\ \hline
{\small Buskerud} & $2.11$ & 1.43 & 1.63 & 1.26 & 1.75 & 1.25 & 2.11 & 1.50\\ \hline
{\small Finnmark} & $2.90$ & 1.63 & 2.69 & 1.63 & 2.94 & 1.61 & 2.90 & 1.81\\ \hline
{\small Hedmark}  & $2.12$ & 1.61 & 2.14 & 1.67 & 1.14 & 1.66 & 2.12 & 1.86\\ \hline
{\small Hordaland} & $2.13$ & 1.25 & 2.13 & 1.18 & 1.78 & 1.23 & 1.69 & 1.13\\ \hline
{\small M\o{}re Og Romsdal} & $2.42$ & 1.46 &  1.54 &  1.54 & 2.19 & 1.94 & 2.42 & 1.62\\ \hline
{\small Nordland} & $2.15$ & 2.08 & 2.59 & 1.84 & 1.88 & 1.75 & 2.47 & 2.25\\ \hline
{\small Nord-Tr\o{}ndelag} & $2.93$ & 2.46 & 2.30 & 2.26 & 2.04 & 1.93 & 2.95 & 2.05\\ \hline
{\small Oppland} & $2.26$ & 1.88 & 2.53 & 1.85 & 2.12 & 1.83 & 2.47 & 1.92\\ \hline
{\small Oslo} & 1.53 & 1.26 & 1.90 & 1.37 & 2.55 & 1.40 & 2.05 & 1.44\\ \hline
{\small \O{}stfold} & $1.90$ & 1.81 & 2.12 & 2.08 & 2.37 & 2.19 & 1.60 & 1.95\\ \hline
{\small Rogaland} & $1.82$ & 1.65 & 2.32 & 1.57 & 2.28 & 1.54 & 2.58 & 1.54\\ \hline
{\small Sogn Og Fjordane} & $2.55$ & 1.44 & 2.19 & 1.36 & 2.50 & 1.36 & 2.05 & 1.36\\ \hline
{\small S\o{}r Tr\o{}ndelag} & $1.72$ & 1.75 & 1.51 & 1.55 & 1.79 & 1.42 & 2.01 & 1.49\\ \hline
{\small Telemark} & $2.87$ & 1.35 & 2.54 & 1.35 & 2.39 & 1.44 & 2.60 & 1.47\\ \hline
{\small Troms} & $2.11$ & 1.45 & 1.23 & 1.43 & 2.60 & 1.54 & 2.41 & 1.00\\ \hline
{\small Vest-Agder} & 2.11 & 1.02 & 2.22 & 1.03 & 2.11 & 0.97 & 2.09 & 1.48\\ \hline
{\small Vestfold} & 2.22 & 1.34 & 2.23 & 1.55 & 2.37 & 1.54 & 2.52 & 1.67\\ \hline
{\small NORWAY} & 1.69 & 1.37 & 1.77 & 1.35 & 2.01 & 1.37 & 1.72 & 1.55\\ \hline\hline
{\small Mean (regions)} & 2.23 & 1.54 & 2.16 & 1.52 & 2.15 & 1.52 & 2.27 & 1.59\\ \hline
{\small Dispersion (regions)} & 0.37 & 0.34 & 0.43 & 0.31 & 0.40 & 0.30 & 0.38 & 0.32\\ \hline
\end{tabular}
\label{tab1}
\end{table}

The mean Pareto exponent $\alpha $ over a four-year period for Norway is
$2.3\pm 0.4$ for income and $1.5\pm 0.3$ for wealth. Note that the
Pareto exponent for income is significantly larger than the Pareto
exponent for wealth. This is the most reliable result thus far obtained
for income and wealth analysis \cite{CG_2005a, CGK_2009, DY_2001a,
  JKP_2012, L_1998, L_2003, NS_2007, YR_2009, JDK_2015, JK_2013b,
  SPW_2004b, CMG_2006, CG_2005b, P_1897, ASF_2003,
  ASNOTT_2000,FSAKA_2003,S_2001,MAH_2004,
  RHCR_2006,S_2006,KY_2004,DNS_2012, JK_2010,KBLMS_2007,LS_1997}. In
addition, the Pareto exponents for income fluctuate in time more than
Pareto exponents for wealth. This means that wealth inequality is more
difficult to change than income inequality.  Note that the results
obtained thus far by authors not using data supplied by tax authorities
are unsystematic and approximate.

Using our comparative analysis we find that for separate regions in
Norway the Pareto exponents for wealth are almost always smaller than
the corresponding Pareto exponents for income. This means that wealth
inequality is higher than income inequality, i.e., the lower the Pareto
exponent, the higher the inequality. This is because income has no
accumulation effect across the generations that acts according to the
preferential choice rule, ``the rich become richer.'' Income inequality
is strongly affected by access to skills and higher education and is
lowered by taxes on income, but wealth accumulation is a long-term
process and is less burdened by taxes, i.e., a cadastral tax or
inheritance tax does not significantly reduce wealth inequality (see
\cite{P_2014}). Although one possible solution to this situation would
be to introduce an annual tax on wealth, social and political factors
make this change difficult \cite{P_2014}.

Using high quality empirical data from the Norwegian tax office, we have
analyzed income and wealth of the richest Norwegian individuals. We find
that income and wealth inequality should be analyzed separately because
they are driven by different factors \cite{P_2014}. In addition, we
confirm that the distribution of top income and wealth is subject to the
Pareto law.

We are aware that the analysis of income and wealth is a research area
for which much has yet to be accomplished. We hope that our paper
contributes some insight into the topic from a physicist's point of
view. The proposed approach is a link between economics and econophysics
and shows that the economic models describing the relationship between
income and wealth can be supported by modelling based on methods used by
physicists.

\end{document}